\documentclass[%
 reprint,
superscriptaddress,
nofootinbib,
 amsmath,amssymb,
 aps,
pra,
]{revtex4-2}
\usepackage{tikz}
\usepackage{graphicx}
\usepackage{dcolumn}
\usepackage{bm}
\usepackage{comment}
\usepackage{notes2bib}
\usepackage{soul}
\usepackage{url}
\bibnotesetup{
note-name = ,
use-sort-key = false
}

\begin{document}


\title{Collective and nonlinear structure of wind power correlations}

\author{Samy E. Lakhal}
\email{samy.lakhal@oist.jp}

 \affiliation{Nonlinear and Non-equilibrium Physics Unit, OIST Graduate University, Onna 904 0495,Japan}

\author{J. E. Sardonia}

\affiliation{Exus Renewables North America, Pittsburgh, Pennsylvania, USA}

\author{M. M. Bandi}
\email{bandi@oist.jp}

 \affiliation{Nonlinear and Non-equilibrium Physics Unit, OIST Graduate University, Onna 904 0495,Japan}

\date{\today}

\begin{abstract}
We describe the correlation structure of wind power fluctuations in a farm of 80 turbines, sampled over 5 years. We report the presence of universal, collective, and nonlinear correlations, responsible for the excess persistency and intermittency of farm-aggregated power output. A first cross-correlation analysis of turbine production reveals a dynamical scaling transition (\textit{à la} Family-Vicszek) from local decoherence to large-scale turbulence-driven scaling, and responsible for the geographical smoothing effect, previously reported beyond farm scale~\cite{bandi2017spectrum}. A second bivariate analysis shows the long-range correlation of non-Gaussian features, responsible for their  amplification in total farm output.  These findings provide important insights into wind energy, whose variability directly results from atmospheric turbulent forcing. By characterizing the multivariate and intermittent structure of fluctuations, our results advance the current understanding of wind power variability and provide valuable priors for grid management and storage optimization, thereby supporting improved integration of wind energy into modern power systems.
\end{abstract}

\maketitle


\section{Introduction \label{sec:Introduction}}
Renewables occupy a growing share of the global energy portfolio. The main challenge of their exploitation is their variability and poor controllability, direct consequences of the fluctuating nature of resources like wind ~\cite{mitigation2011ipcc,peinke2004turbulence,manwell2010wind} and solar radiation~\cite{KlimaAptBandi2018,BelBandi2019,BelBandi2024}. As a result, renewables constitute a source of risk --- both local and global ---  on infrastructure integrity~\cite{menck2014dead,rohden2012self,apt2014variable,Bel2016,tavner2011correlation}, adaptability to demand~\cite{schmietendorf2017impact}, and market pricing regulation~\cite{HU2021105159,BRANCUCCIMARTINEZANIDO2016474}.

To ensure the safe and efficient integration of these sustainable energy sources, a deep understanding of generation-side fluctuations ~\cite{veers2019grand,van2016long,Peinke2024Reflets} becomes essential. For wind energy in particular, turbine power output $P(t)$ fluctuations mirror the both scale-invariant and intermittent nature of atmospheric wind turbulence~\cite{APT2007369,bandi2016variability,bandi2017spectrum,bossuyt2017wind,nanahara2004smoothing,haehne2018footprint,anvari2016short,liu2013cascade,beck2005atmospheric,muzy2010intermittency,baile2010spatial,calif2013,calif2014,ali2019-2}, roughly ranging from turbine reaction time ($\tau_R\sim 10$s) to day/night injection cycles ($T\sim 24$h). 

Beyond single turbine production, the larger scale impact of a wind farm on the energy grid is evaluated by its total output production ($P_\mathrm{tot} = \sum_i P_i$). Direct farm level data analysis has shown the emergence of smoothing effect under geographical averaging, at farm level~\cite{nanahara2004smoothing} or beyond~\cite{bandi2017spectrum}, and its translation into steeper spectral decay. This apparent advantage ---more persistent fluctuations might intuitively suggesting improved short-term predictability--- is rapidly mitigated by the statistical properties of both individual and aggregated signals. In particular, spatially averaged power outputs retain and even amplify the non-Gaussian features observed at the single-turbine level (see distributions in Ref.~\cite{milan2013turbulent}), whose heavy-tailed nature can give rise to potentially damaging extreme events. Understanding these robust statistics across scales is thus a key requirement in the global and sustainable integration of wind-energy.

This article reports an analysis of the statistical structure of wind power production in a wind farm of 80 turbines, that spans over $20\mathrm{km}$.  To do so, the wind speed $v_i(t) = \|\boldsymbol{v}_i\|$ and wind power $P_i(t)$ of each individual turbine were sampled every 10 minutes ---following International Energy Commission standard 61400-12~\cite{IEC61400-12-1}--- for over 5 years. The turbines, of rotor hub height $80\mathrm{m}$ and $90\mathrm{m}$ and diameter $116 \mathrm{m}$ are organized in rows with an average separation of $365$m, and positioned on slightly elevated rolling crest lines of maximum height difference $\Delta = 150$m.  Each rows are facing the direction of the most prevailing wind directions from the south-southwest origin, and are separated by a minimum distance of $\sim 1\mathrm{km}$, beyond the near wake of turbines~\cite{amiri2024review,ali2016,bachant2015characterising,ali2019}.

Technical information of the wind farm considered is reported in Supplemental Material I.~\bibnote[Supplemental]{See Supplemental Material [url] for: (I) detailed turbine specifications; (II) informations on the cleaning procedure applied to wind-power time series; (III) a comparison of scaling properties across turbines; (IV) a comparison of standardized structure functions for individual and aggregated wind power; and (V) a copula-based analysis of wind-power fluctuations.}. In Sec.~\ref{sec:WindStats}, a first discussion on the turbulent nature of atmospheric winds (\ref{subsec:IIA}) motivates the description of turbine statistics through increments (\ref{subsec:IIB}) and multifractal analysis (\ref{subsec:IIC}). In Sec.~\ref{sec:CorrelationStructure}, these characterizations are extended at farm scale to understand the persistence of extreme fluctuations over farm aggregation~(\ref{subsec:IIIA}): a cross-correlation analysis reveals the coherent structure of wind induced turbine fluctuations, and its impact over aggregated production~(\ref{subsec:IIIB}). Ultimately, we describe the spatial persistency of extreme events, among and beyond the farm~(\ref{subsec:IIIC}).

\section{From atmospheric turbulence to wind turbine statistics \label{sec:WindStats}}
\subsection{Atmospheric wind turbulence \label{subsec:IIA}}
\emph{Increments statistics.---} Turbulent flows, commonly characterized through their velocity, pressure, or temperature fields, are known to exhibit multiscaling properties~\cite{Kolmogorov1941,frisch1995turbulence,friedrich2022superstatistical,CASTAING1990177,chevillard2005intermittency,pereira2018multifractal}.
Formally, for a multiscaling field \(X(\boldsymbol{r},t)\), the increments
\(\delta_{\boldsymbol{\ell},\tau} X = X(\boldsymbol{r}+\boldsymbol{\ell},t+\tau) - X(\boldsymbol{r},t)\)
exhibit anomalously scaling absolute moments, namely the structure functions
\[S_q(\boldsymbol{\ell},\tau) = \big\langle \lvert \delta_{\boldsymbol{\ell},\tau} X \rvert^q \big\rangle.\]
In particular, temporal increments satisfy
\begin{equation}
    S_q^X(\tau)= \langle \lvert \delta_{\tau} X \rvert^q \rangle \propto \tau^{\zeta_q}, \qquad \tau_\eta \ll \tau \ll T,
    \label{eq:SpaceTimeScale}
\end{equation}
where the scaling exponent spectrum \(\zeta_q\) characterizes the self-similar properties of \(X(t)\) from the energy-injection timescale \(T\) down to the viscous microscale \(\tau_\eta\). Here, the generalized \textit{Hurst} (or roughness) exponent spectrum \(H_q = \zeta_q/q\) quantifies the temporal persistence of fluctuations, from typical events (\(q \ll 2\)) to extreme ones (\(q \gg 2\)). While a monofractal process with \(H_q = H\) exhibits scale-invariant jump statistics, as in fractional Brownian motion, a multifractal process with \(H_q \neq H\) displays continuously evolving statistics across scales~\cite{bacry2001multifractal,Lakhal2025}. For example, a decreasing spectrum \(H_q\) corresponds to the presence of increasingly heavy-tailed distributions at small scales, a property commonly referred as \textit{small-scale intermittency}.\\

\emph{Inertial range of atmospheric turbulence.---} In atmospheric turbulence, the low kinematic viscosity of air implies velocity fluctuations extending down to very short timescales, of the order of the Kolmogorov timescale \(\tau_\eta \sim 10\,\mu\mathrm{s}\). Here, however, we focus on the large-scale forcing time \(T\), beyond which fluctuations decorrelate $(S_q(\tau \gg T) \to \mathrm{const})$, and contrast two physically distinct interpretations for its origin.

A first estimation assumes that turbulent structures are generated by the vertical shear of the mean wind, with representative speed \(v_0 = \langle v \rangle \sim 8\,\text{m.s}^{-1}\), across the height of the atmospheric boundary layer, of characteristic height \(L \sim 1\,\mathrm{km}\). Assuming the turbulence to be locally produced by mean flow gradients, one can compute a characteristic timescale \(T\approx {L}/{ v_0 }\sim 125\,\mathrm{s}\).
However, this timescale lies well below the upper correlation ranges
reported in atmospheric measurements~\cite{bandi2017spectrum,APT2007369}.

Instead, a second sweeping-based interpretation associates the decorrelation time with the diurnal cycle, \(T \simeq 24\,\mathrm{h}\), as directly observed in wind-speed records~\cite{bandi2017spectrum}. In this picture, turbulent structures are advected horizontally by the largest energy-containing motions at a root-mean-square velocity \(u_{\mathrm{RMS}} = \sqrt{\langle \delta_T v^2 \rangle}\sim 5\,\text{m.s}^{-1}\)~\cite{he2011kraichnan,kraichnan1964kolmogorov}. The associated sweeping length, \(L = u_{\mathrm{RMS}} T \approx 400\text{--}500\,\mathrm{km},\) defines a mesoscale range consistent with meteorological measurements~\cite{bandi2017spectrum,baile2010spatial,muzy2010intermittency}. This large-scale sweeping picture is further supported by the observation of spatial and temporal Kolmogorov scaling~(\(\zeta_3 = 1\))~\cite{Kolmogorov1941,frisch1995turbulence}, across various altitudes~\cite{nastrom1984kinetic,bandi2017spectrum,baile2010spatial,muzy2010intermittency}.\\

\emph{Space-time structure.---} A joint space–time description of turbulent statistics provides a powerful experimental framework for characterizing correlations over the full spatiotemporal domain. Taking cue from prior turbulence experiments and numerical studies~\cite{he2011kraichnan,he2006elliptic}, it was shown that, for isotropic turbulence under linear advection, iso-$S_q$ contours in space–time could be convincingly approximated by elliptical shapes (see, however, Ref.~\cite{he2011kraichnan} for convective corrections). This leads to an effective parametrization that incorporates both advection and sweeping effects,
\begin{equation}
S_q(\boldsymbol{\ell},\tau) \propto
\bigg((\boldsymbol{\ell}-\boldsymbol{v}_0 \tau)^2 + u_{\mathrm{RMS}}^2 \tau^2 \bigg)^{\zeta_q/2},
\label{eq:EllipticApproximation}
\end{equation}
and which interpolates between the Galilean-invariant Taylor frozen-turbulence limit ($v_0\gg u_\mathrm{RMS}$) and the continual renewal of turbulent structures described by the Kraichnan–Tennekes random-sweeping hypothesis ($v_0\ll u_\mathrm{RMS}$)~\cite{kraichnan1964kolmogorov,tennekes1975eulerian}.

While the spatial scaling of upper-atmospheric winds is well known to follow Kolmogorov’s K41 $k^{-5/3}$ law~\cite{nastrom1984kinetic}, the ABL is expected to exhibit different and nontrivial spatial features. First, the turbulent organization of the ABL~\cite{Liu2021Universal} undergoes strong diurnal modulation, ranging from daytime buoyancy-driven convection to nighttime stable stratification at night. Second, viscous effects induced by obstacles—such as topographic variations~\cite{druault2022spatial}—introduce local retardation and disrupt the steadiness of wind flows. Third, advection may remain observable between turbine rows in the downwind direction (south–southwest in the dataset analyzed here). However, taking an inter-row distance of $ 1\mathrm{km}$ and representative wind speed $v_0=8 \mathrm{m.s}^{-1}$  yields an advection time $\tau_\mathrm{advection} = 1000\mathrm{m}/8\mathrm{m.s^{-1}} = 125\mathrm{s}$  well below our sampling resolution $\Delta t = 10\mathrm{min}$. For larger separations, these causal structures should be further dampened by interactions with terrain and other obstacles.
We anticipate both effects of Eq.~\ref{eq:EllipticApproximation} to mix into a single effective sweeping contribution of leading order:
\begin{equation}
    S_q(\boldsymbol{\ell} ,\tau) \propto \left(\tau^2 +\tau^2_{\boldsymbol{\ell}}\right)^{\zeta_q/2},
\label{eq:EffectiveSweeping}
\end{equation}
where the \textit{coherence time} \(\tau_{\boldsymbol{\ell}}\) sets the characteristic sweeping time over distance \(\ell\) below which fluctuations should decorrelate (\(S_q(\tau \ll \tau_{\boldsymbol{\ell}}) \to \mathrm{const}\)) and beyond which multiscaling should be retrieved. Furthermore, we expect \(\tau_{\boldsymbol{\ell}}\) to grow sublinearly with \(\ell\), due to the local retardation of small smaller-scale with the environment, a conjecture that will be examined below.

\subsection{Extreme power fluctuations\label{subsec:IIB}}

\begin{figure}
    \centering
    \includegraphics[width=1.\linewidth]{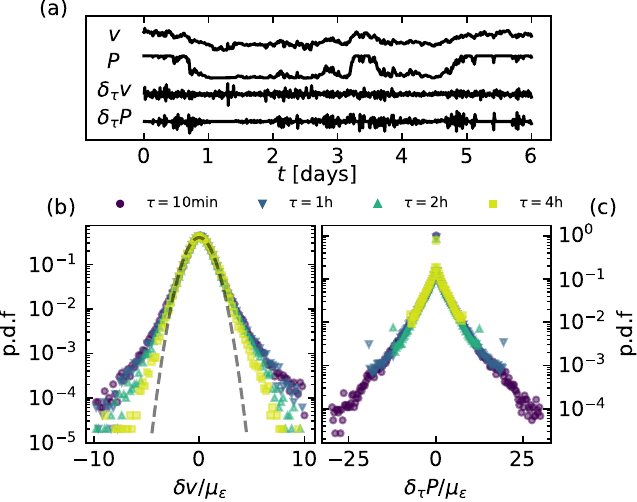}
    \caption{Jumps statistics of wind speed $v$ and wind power $P$. (a) Weekly evolution of wind speed $v$, power output $P$  and their jumps $\delta_\tau v$ and $\delta_\tau P$ ($\tau = 10$min). (b) and (c) Probability density functions (p.d.f) of jumps 
    ($\delta_\tau v,\delta_\tau P$) for $\tau = 10\text{min},1\text{h},2\text{h},4\text{h}$. To avoid the bias introduced by cutoff effects (especially for $P$), the core of p.d.f.s has been collapsed by rescaling jumps the ratio between empirical and Gaussian low order ($\epsilon \ll 1$) moments: $\mu_\epsilon =[{\langle |\delta_\tau X|^{\epsilon} \rangle \, \sqrt{\pi}}{2^{-\epsilon/2}/\Gamma\!\left(\tfrac{1+\epsilon}{2}\right)}]^{\!1/\epsilon}$.}
    \label{fig:Time_PDF}
\end{figure}

Wind turbines convert a nearly constant fraction of the incoming wind power \(P_{\text{wind}}(t)\) into electrical power \(P(t)\). The wind power, associated with the transport at speed \(v\) of a cylindrical volume of air carrying kinetic energy \(E_c \propto v^2\), leads to the well-established cubic scaling \(P \propto v^3\)~\cite{bandi2016variability}. Figure~\ref{fig:Time_PDF}(a) illustrates the close correspondence between the temporal fluctuations of \(v(t)\) and \(P(t)\) for a representative turbine in the farm. Deviations from the cubic law arise outside the operational wind-speed range of the turbines, bounded by the cut-in (lower) and rated (upper) speeds~\cite{bandi2016variability}, beyond which the electrical power saturates. The overall turbine response curves \(P = f(v)\) for the two turbine models considered in this study are reported in Supplemental Material~I.~\bibnotemark[Supplemental].

Similar to standard turbulence analysis, we describe the statistical structure of wind speed and wind power by computing their increments $\delta_\tau v$ and $\delta_\tau P$. We see in Fig.~\ref{fig:Time_PDF}(a) that such finite difference renders the signal stationary, and now evolves around a clearly defined average. The envelopes of the increments, \((|\delta_\tau v|, |\delta_\tau P|)\), themselves fluctuate over larger correlation scales, displaying alternating phases of weak local activity periods and intermittent bursts~\cite{Vernede2015,Lakhal2025}. These stationary yet intermittent variations give rise to time-aggregated non-Gaussian jump statistics, as illustrated in Fig.~\ref{fig:Time_PDF}(b) and (c), where normalized probability density functions (see caption for standardization rule) exhibit increasingly heavy tails at small \(\tau\), reflecting small-scale intermittency. For \(\delta_\tau v\), the jump statistics progressively converge toward a Gaussian distribution as \(\tau\) increases. In contrast, for \(\delta_\tau P\) the turbine response function \(P=f(v)\) imposes a hard cutoff, confining the power to the interval \([0,P_{\text{sat}}]\). This constraint leads to a condensation of probability near \(\delta_\tau P=\pm P_{\text{sat}}\) for the largest increments, and near \(\delta_\tau P=0\) during shutdown or rated operating regimes. As shown in Supplemental Material~II.~\bibnotemark[Supplemental], defining a corrected power signal \(P_c\), obtained by removing these saturating regimes, is sufficient to retrieve the statistics of \(v^3\), except for extreme fat-tail events that cannot be reproduced by such filtering.

\subsection{Multifractal analysis \label{subsec:IIC}} 
Following seminal contributions to turbulence~\cite{frisch1995turbulence,Kolmogorov1941,mandelbrot1989multifractal} and more recent analytical developments~\cite{bacry2001multifractal,mandelbrot2005possible,Lakhal2025,chevillard2020multifractal}, it is now well established that the non-Gaussian statistics of turbulent flows can be efficiently described within the multifractal formalism. In the following, we investigate the scaling properties of our wind farm dataset.

\textit{From velocity to speed and power.---} We expect the intermittent, non-Gaussian statistics of  \(v\) and \(P\) to originate from the multifractal nature of the underlying turbulent wind velocity field \(\boldsymbol{v}\)~\cite{frisch1995turbulence}, as encoded in the scaling properties of its increments. To first order, the increments of speed and power can be approximated as
\begin{equation}
    \delta_\tau v
    = \delta_\tau \boldsymbol{v}\cdot \boldsymbol{e}_v,
    \qquad
    \delta_\tau P
    = v^2\,(\delta_\tau \boldsymbol{v}\cdot \boldsymbol{e}_v)
    = v^2\,\delta_\tau v,
    \label{eq:FirstOrder}
\end{equation}
where \(\boldsymbol{e}_v\) denotes the unit vector in the direction of the mean wind velocity \(\boldsymbol{v}\). In both expressions, velocity increments are effectively projected onto the dominant flow direction. While identical scaling properties are therefore expected for \(\boldsymbol{v}\) and \(v\), the multiplicative factor \(v^2\) in the power increment introduces additional variability, leading to potential deviations in the statistics of \(P\), as discussed below.

For real wind turbines, however, the electrical power is bounded by the rated value \(P_{\mathrm{sat}}\). In this regime, the effective scaling of power increments, \(\delta_\tau P \sim P^{2/3}\,\delta_\tau v\), is expected to closely follow that of the wind speed, yielding
\[ S_q^{\boldsymbol{v}} \sim S_q^{v} \sim S_q^{P} \sim \tau^{\zeta_q}. \]
The limiting case of ideal turbines, for which \(P_{\mathrm{sat}} \to +\infty\), will be discussed separately.\\

\begin{table}
\caption{\label{tab:MultifractalParameters} \textcolor{black}{Statistics of wind speed $v$, individual turbine power $P$, and total farm power $P_{\mathrm{tot}}$. For ($v,P$) most extreme values among turbines were reported.}}
\begin{tabular}{c|r|r|r}
\hline
\hline
 X &$v$&$P$&$P_{\mathrm{tot}}$\\
\hline
{$\langle X \rangle$}& $8.0 - 9.0$ m.s$^{-1}$ & $1.09- 1.36$ MW & 100.2  MW\\
{$\sqrt{{\langle \delta_{T} X^2 \rangle}}$}& $4.4- 5.3$ m.s$^{-1}$ & $1.12-1.28 $ {MW}& 91.6 {MW}\\
{$\sqrt{{\langle \delta_{T} X^2 \rangle}}/\langle X\rangle$}& $0.54 - 0.64$ & $0.90-1.04$ & 0.93\\
{$\zeta_3$}&$1.05\pm 0.03$&$0.95\pm 0.03$& $1.10\pm 0.04$\\
\hline
\hline
\end{tabular}
\end{table}

\begin{figure}[b]
    \centering
    \includegraphics[width=1.\linewidth]{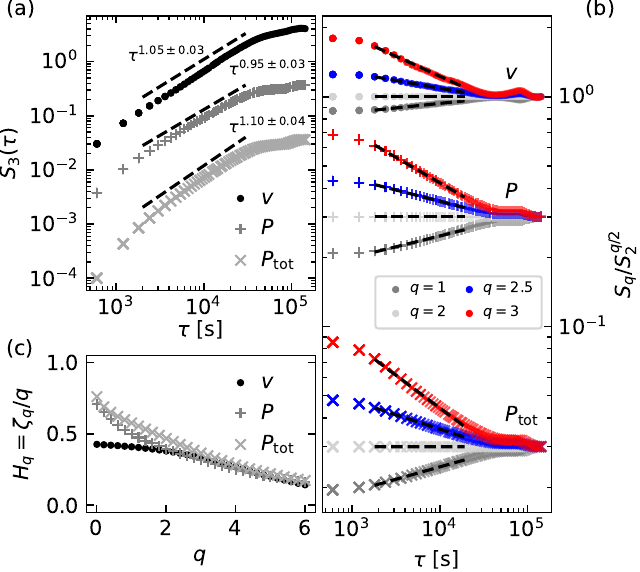}
    \caption{Intermittency analysis of wind speed $v$ ($\circ$), single turbine power output $P$ ($+$) and total farm production $P_\mathrm{tot} = \sum_i P_i$ ($\times$). (a) $S_3(\tau)$ vs $\tau$. (b) $S_q(\tau)$ rescaled by $(S_2(\tau))^q $ for $q =1,2,2.5,3$, fits for $30 \text{min} \leq\tau\leq 5\text{h}$ in black dotted lines. (c) Scaling exponent spectrum $H_q = \zeta_q/q$. }
    \label{fig:Intermittency}
\end{figure}

\textit{Kolmogorov scaling.---} Figure~\ref{fig:Intermittency}(a) shows the third-order structure functions \(S_3(\tau)\) of both \(v\) and \(P\) for a representative turbine, which display clear power-law scaling up to \(\tau \sim 12\,\mathrm{h}\), in accordance with the diurnal forcing timescale. A fit performed over approximately one decade, \(30\,\mathrm{min} \leq \tau \leq 5\,\mathrm{h}\) ---the smallest scales being typically affected by windowing procedures--- yields \(S_3(\tau) \propto \tau^{\zeta_3}\) with \(\zeta_3 \simeq 1\). We show in Supplemental Material III.~\bibnotemark[Supplemental] that this scaling behavior is observed for all turbines in the farm and is consistent with previous measurements reported in Ref.~\cite{bandi2017spectrum}, down to the turbine response time \(\tau_R \approx 30\,\mathrm{s}\). It corresponds to the Kolmogorov (K41)~\cite{Kolmogorov1941} prediction for a direct turbulent cascade, as was similarly reported in Dutch and Corsican meteorological observations~\cite{baile2010spatial,muzy2010intermittency}.

This result indicates that, within a real wind-farm configuration, the temporal autocorrelation properties of wind-speed and power fluctuations are essentially identical across turbines, although this does not imply their statistical independence. This behavior contrasts with results from idealized experiments and numerical simulations, where regularly designed farms with periodic turbine placement, linearly advected turbulence, and identical rotor heights could imprint modification on individual spectral behavior~\cite{liu2017towards,chatterjee2018contribution,SINGH2025106537,tobin2018turbulence}. Here, the subjection of wind turbines to real meteorological and topographical conditions suppresses such effects.\\

\textit{Multifractal analysis.---} The full scaling spectrum \(\zeta_q\) is obtained by combining two independent measures: \(\zeta_3\), extracted from the fit shown in Fig.~\ref{fig:Intermittency}(a), and the deviations \([\zeta_q - q\zeta_2/2]\), inferred from the ratios \(S_q/S_2^{q/2}\) displayed in Fig.~\ref{fig:Intermittency}(b). That last observable extends the classical flatness (\(q=4\)) commonly used in turbulence studies into a \textit{generalized flatness} measure. As mentionned in previous works~\cite{santucci2007statistics}, fitting ratios of structure functions effectively enlarges the inertial range and improves the robustness of scaling exponents. Consistent results are also obtained using an Extended Self-Similarity (ESS) analysis~\cite{benzi1993extended,milan2013turbulent}, however without enforcing the constraint \(\zeta_3 = 1\). The resulting generalized Hurst exponent spectrum, \(H_q = \zeta_q/q\), is reported in Fig.~\ref{fig:Intermittency}(c), and its behavior is discussed below.

First, in regard of previously established works~\cite{bandi2017spectrum}, the same exponents $H_2 \approx 0.4$ were retrieved, in full correspondance with other energy spectrum estimates~\cite{APT2007369,bandi2017spectrum}. This allows us to discuss the relative position of exponents at lower and higher order. For \(q < 2\), moments primarily describe the scale invariance of small-amplitude fluctuations, at the core of increments distributions. Turbine shutdown and saturated regimes produce persistent extended periods of low amplitude jumps \(\delta_\tau P \approx 0\), resulting in larger roughness exponents \(H_q\) for \(P\) compared to \(v\). For \(q>2\), the moments become sensitive to larger and extreme fluctuations. However, the mean-reverting statistics of \(v\) induce an effect analogous to the bounded structure of \(P\), leading to similar exponents.
\begin{figure}[b]
    \centering
    \includegraphics[width=1.\linewidth]{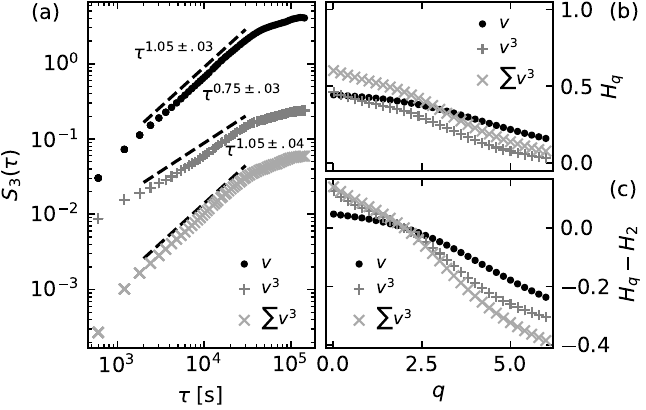}
    \caption{Intermittency analysis of wind speed $v$ ($\circ$), $v^3$ ($+$) and $\sum v^3$ ($\times$). (a) Structure functions $S_3$ VS $\tau$. (b) Generalized Hurst exponents $H_q = \zeta_q/q$ vs $q$, fitted as in Fig.~\ref{fig:Intermittency} (c) $H_q-H_2$ vs $q$.}
    \label{fig:IntermittencyV3}
\end{figure}\\

\textit{Extension to ideal turbines.---} 
The scaling properties of the measured power \(P\) are strongly influenced by the response function of turbines. Therefore, we extend the analysis to ideal turbines with infinite rated power (\(P_{\mathrm{sat}} \to +\infty\)) by directly examining \(v^3\) and, the corresponding ideal farm output \(\sum_i v_i^3\). Figure~\ref{fig:IntermittencyV3}(a) reveals a clear scaling behavior for \(v^3\), with a roughness exponent \(H_3 \simeq 0.25\), smaller than Kolmogorov predictions (\(H_3 = \zeta_3/3 \simeq 1/3\)), and identically observed for $P_c$ (not shown here).

We attribute this effect to the auto-regressive properties of wind speed, reported in 
Ref.~\cite{bandi2016variability} and which results in positive correlations between wind amplitude $v$ and increments amplitude $
|\delta_\tau v|$, i.e. strong mean winds yield stronger wind fluctuations. As a result, the increments of exponentiated velocity should scale as a higher order power law of wind speed increments:  \[\delta_\tau (v^m)  \approx v^{m-1}\,\delta_\tau v\sim (\delta_\tau v)^{\alpha_m},\] with $\alpha_m>1$, increasing with $m$. In consequence, wind power increments ($m=3$ for ideal turbines) will be shifted, $q\to \alpha_m q$, and  yield the decrease in $H_2$ reported for $v^3$ in  Fig.~\ref{fig:IntermittencyV3}(b). We found, by collapsing the exponent spectra of $v^2$ and $v^3$ onto $v$ that $\alpha_2 \approx 1.3$ and $\alpha_3 \approx 1.5$.

\section{Correlation structure of wind farm production \label{sec:CorrelationStructure}}
\subsection{Statistics of aggregated turbines \label{subsec:IIIA}} 

\begin{figure}
    \centering
    \includegraphics[width=.7\linewidth]{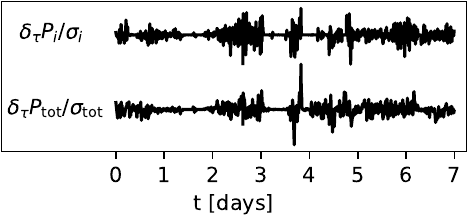}
    \caption{Wind power increments $\delta_\tau P$ ($\tau = 10\text{min}$) of a single turbine (top) and of the total farm (bottom), sampled over a week . Each time series  was normalized by its empirical deviation.}
    \label{fig:SingleTotalIncrements}
\end{figure}

From the standpoint of wind farm operators, the total farm production $P_\mathrm{tot} = \sum_{i} P_i$ is an observable of interest for power transmission and grid design. Although one would expect the summation of production over turbines to yield smoother Gaussian statistics and fewer extreme events, we will see below that this intuition does not hold. A preliminary result may be that the daily deviation to mean ratio only weakly decays from $P_i$ to $P_{\mathrm{tot}}$ (see Tab.~\ref{tab:MultifractalParameters}), and remains significantly larger than the discount predicted by the Central Limit Theorem (CLT) $N^{-1/2}\sim 0.1$. We expect such behavior to result from the presence of temporal and spatial correlations among and between turbines.  That intuition is confirmed in Fig.~\ref{fig:SingleTotalIncrements}, where a qualitative comparison of single and farm aggregated turbine output statistics evidences strong similarities, which we now investigate quantitatively.

As shown in Fig.~\ref{fig:Intermittency}(a), the aggregated farm production \(P_{\mathrm{tot}}\) reproduces the scale-invariant behavior observed for individual turbine power \(P\), with however modified scaling exponents. Timescales shorter than the sweeping time across the farm, \( \tau_D = D/u_{\mathrm{RMS}} \sim 2\,\mathrm{h}, \) are expected to be affected by spatial aggregation. In practice, the relaxation toward decorrelation and the presence of retarded structures extend these modifications up to \(2\text{--}3\,\tau_D\); see Supplemental Material~IV.~\bibnotemark[Supplemental] for a direct comparison between the scaling behavior of  \(P\) and \(P_{\mathrm{tot}}\). The resulting generalized Hurst exponents \(H_q\), shown in Fig.~\ref{fig:Intermittency}(c), are systematically larger for \(P_{\mathrm{tot}}\) at low orders (\(q<2\)), while they coincide with those of \(P\) for higher orders (\(q>2\)). This respectively indicates the emergence of a more spread distribution core, as well as the conservation of non-Gaussian tail properties over aggregation.  Note that the increase of \(H_2\) indicates the emergence of more persistent time correlations.

While the second-order exponent \(H_2\) primarily characterizes the correlation structure of the time series—persistent (antipersistent) dynamics corresponding to \(H_2>1/2\) (\(H_2<1/2\)) ---the scale dependence of standardized jump statistics is more naturally captured by the relative roughness exponent \(H_q-H_2\), directly fitted from generalized flatness functions. In the idealized limit \(P_{\mathrm{sat}}=+\infty\), Fig.~\ref{fig:IntermittencyV3}(c) shows that low-order relative exponents remain essentially unchanged, whereas high-order ones decay sharply: the core of the increment distributions is preserved, while extreme events, unbounded here, constructively add up to extend the tail of aggregated production.

\subsection{Cross-structure analysis \label{subsec:IIIB}}

The multifractal analysis introduced above confirms the presence of cross-correlations among turbines, and which we now directly describe. As shown previously and further described in Supplemental~I.~\bibnotemark[Supplemental], the statistics of $P$ were found to be similar to $v^3$ within the operating regime $P\propto v^3$. To describe the wind-field structure dictating turbine production, we then directly use wind speeds $(v_i)$, standardized by their deviations and mean values. We describe pairwise correlations by using cross-structure functions: \[S_q^{ij}(\tau) = \langle |v_i(t+\tau)-v_j(t)|^q \rangle.\] We focus on the second order case $q=2$ to retrieve the a direct equivalence with cross-spectrum  analysis $\overline{\mathcal{F}[{v}_i]}\mathcal{F}[{v}_j]$~\cite{liu2024turbulence}. Note that replacing $v$ by $v^3$, or cleaned turbine output $P_c$ yields identical conclusions in the following.\\

\begin{figure}
    \centering
    \includegraphics[width=.9\linewidth]{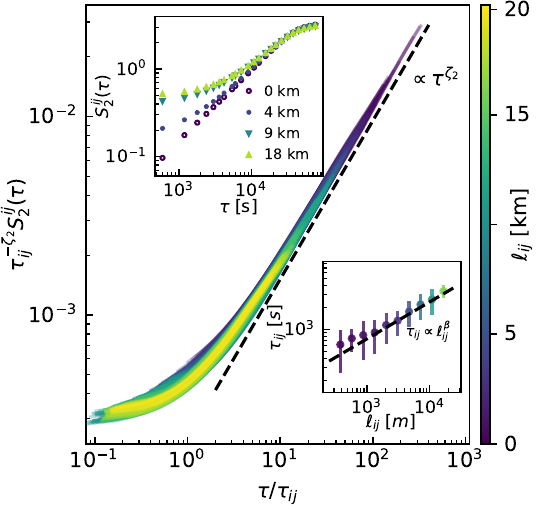}
\caption{Cross-structure analysis of wind turbines. (Upper inset) $S_2^{ij}$ \textit{v.s} $\tau$ for neighbouring turbines, with pair turbine distance $\ell_{ij}$ in legend. (Main figure)  Rescaled $S_2^{ij}$ \textit{v.s} $\tau$, fitted and collapsed using Eq.~\eqref{eq:FitCurve}. Scaling asymptote in black with $\zeta_2 = 0.8$\ . (Lower inset) Fitted coherence times ${\tau}_{ij}$ as functions of $\ell_{ij}$, fitted power law scaling in black with $\beta = 0.60 \pm 0.15$.}
    \label{fig:SpatialIntermittency}
\end{figure}

\textit{Cross-structure analysis.---} Figure.~\ref{fig:SpatialIntermittency} (upper inset) shows $S_2^{ij}(\tau)$ for 4 turbines separated by distances $\ell_{ij}$ ---$\ell_{ij} = 0$ corresponding to $i=j$. Two scaling regimes, postulated in Eq.~\eqref{eq:EffectiveSweeping}, appear: \textit{(i)} for low $\tau$, $S_2^{ij}$ flattens as turbulence decorrelates at short time scales. \textit{(ii)} Beyond this short time scale, signal coherence~\cite{liu2024turbulence} is retrieved as an $\ell_{ij}$-dependent transition towards scale invariance,~$S_2^v\propto \tau^{\zeta_2}$.

This asymptotic behavior being identical to the effective sweeping predicted in Eq.~\eqref{eq:EffectiveSweeping}, we decided in Fig.~\ref{fig:SpatialIntermittency}(lower inset) to fit and display, for all $(i,j)$, an \textit{effective coherence time} parameter $\tau_{ij}$ defined from:

\begin{equation}
    S_2^{ij}(\tau) \propto \left( \tau^2+\tau_{ij}^2 \right)^{\zeta_2/2}.
    \label{eq:FitCurve}
\end{equation}

We found ${\tau}_{ij}$ to evolve from a few minutes to the previously estimated sweeping time over the farm  $\tau_D = 2\mathrm{h}$. Estimation of sweeping speed $u_{ij}=\ell_{ij}/\tau_{ij}$ for neighboring turbines ($u_{ij}= 500\text{m}/300\text{s} = 1.3\mathrm{m.s}^{-1}$) indicates that the decoherence regime over short distances is dictated by slower, more retarded sweeping dynamics. This result is confirmed in  Fig.~\ref{fig:SpatialIntermittency} (lower right inset), where a power-law fit $\tau_{ij}\propto \ell_{ij}^{\beta}$  $\beta= 0.60 \pm 0.15$ evidences the sublinear growth of sweeping times over separation $\ell_{ij}$.\\

\textit{Influence of ABL regimes.---} The scattering of measured $\tau_\ell$ has multiple causes: wake patterns, rough topography, and seasonal changes. In particular, we know the the wind-speed correlation structure to be strongly affected by day/night modulations of the ABL. Indeed, during the day, solar heating creates vertical mixing that brings stronger winds from higher altitudes down to the surface, making winds typically stronger and gustier. At night, the ground cools, stabilizing the atmosphere and suppressing vertical mixing, which leads to calmer winds with horizontally stratified structures.
We reproduce the analysis of Fig.~\ref{fig:SpatialIntermittency} by masking $(v_i)$ into day (vertical/convective) and night (horizontal/sweeping) ABL regimes~\cite{Liu2021Universal}. We retrieved $\beta = 0.3 \pm 0.1$ and $\beta = 0.8 \pm 0.1$, corresponding to \textit{(i)} convective turbulence for which $\tau_{ij}$ is expected to saturate~\cite{he2011kraichnan,zhou2011experimental} ($\beta=0$), and horizontally swept/advected turbulence for which $\tau_{ij}\propto\ell_{ij}$ ($\beta=1)$. We expect higher frequency sampling over a few days to yield more refined estimates of $\beta$.\\

\textit{Collapse to master curve.---} In the main plot of Fig.~\ref{fig:SpatialIntermittency}, $S^{ij}_2$ v.s.~$\tau$ curves were rescaled from Eq.~\eqref{eq:FitCurve} using $\zeta_2 = 0.8$. As a result, all dataset collapsed onto a single master curve $x \mapsto f(x)$ defined by:
\[f(x)\sim \left\{\begin{array}{ll}
   1  & \mathrm{for} \ x\ll 1\\
   x^{\zeta_2}  & \mathrm{for} \ x\gg 1,
\end{array}\right.
\label{eq:FamilyVicszek}\]
where $x = \tau/\tau_{ij}$. This form is directly analogous to dynamic scaling, \textit{à la Family-Vicszek} where space-time correlations collapse onto a universal function. The coherence time \(\tau_{ij}\) encapsulates a strong statistical dependence on turbine separation \(\ell_{ij}\), demonstrating that the transient and coherence regimes are universal across the farm.

\textit{From correlations to spectral steepening.---}{ The increase of $P_\text{tot}$'s scaling exponents can now be traced back to the cross-correlation identified previously. For $q=2$, the structure function of total power writes: \[S^{\mathrm{tot}}_2(\tau) = \sum_{i} S_2^{ii}(\tau) +\frac{1}{2}\sum_{i\neq j}\big[S_2^{ij}(\tau) - S_2^{ij}(0)\big],\]
which, in the limit $N\gg 1$, is dominated by the second r.h.s sum involving $N(N-1)\sim N^2$ contributions. These pairwise contributions can be expanded following:
\[
S_2^{ij}(\tau) - S_2^{ij}(0) \propto \tau_{ij}^{\zeta_2}\left[f(\frac{\tau}{\tau_{ij}})-f(0)\right]\sim \left(\frac{\tau}{\tau_{ij}}\right)^{\alpha},
\]
where the exponent $\alpha$ controls $f$'s scaling properties around 0. In particular, the effective sweeping form of Eq.~\eqref{eq:EffectiveSweeping} predicts ballistic scaling with \(\alpha = 2\), whereas the elliptic approximation of Eq.~\eqref{eq:EllipticApproximation}, dominated by advection effects, leads to diffusive scaling with \(\alpha = 1\). We note, however, that the scaling range of the individual cross-structure terms \(S^{ij}_2(\tau) - S^{ij}_2(0)\) is not sufficiently extended in the present dataset to allow for a definitive discrimination between these two regimes.

}\smallskip

\subsection{Extreme fluctuations: excess intermittency and range \label{subsec:IIIC}}

\textit{Copula analysis.---} While the cross-scaling analysis above accounts for the increase of the roughness exponents \(H_q\) under turbine-output aggregation, it does not by itself explain the persistence of non-Gaussian statistics, nor their enhancement, at short timescales (\(\tau \to 0\)). This effect is direclty observable in Fig.~\ref{fig:SingleTotalIncrements} where the relative amplitude of calm to extreme fluctuations appear is visually higher for $P_\mathrm{tot}$ than for a single turbine.

To address this point, we turn to a multivariate description of power increments and analyze the joint probability density functions \(\mathbb{P}(\delta_\tau P_i,\delta_\tau P_j)\), focusing on their dependence on the time lag \(\tau\) and the inter-turbine separation \(\ell_{ij}\). As shown in Supplemental Material~V.~\bibnotemark[Supplemental], the progressive alignment of the iso-\(\mathbb{P}\) contours along the diagonal reveals the buildup of correlations between power fluctuations, as \(\tau\) increases and \(\ell_{ij}\) decreases, corresponding respectively to temporal integration and turbine separation.

\begin{figure}[t]
    \centering
    \includegraphics[width=1.\linewidth]{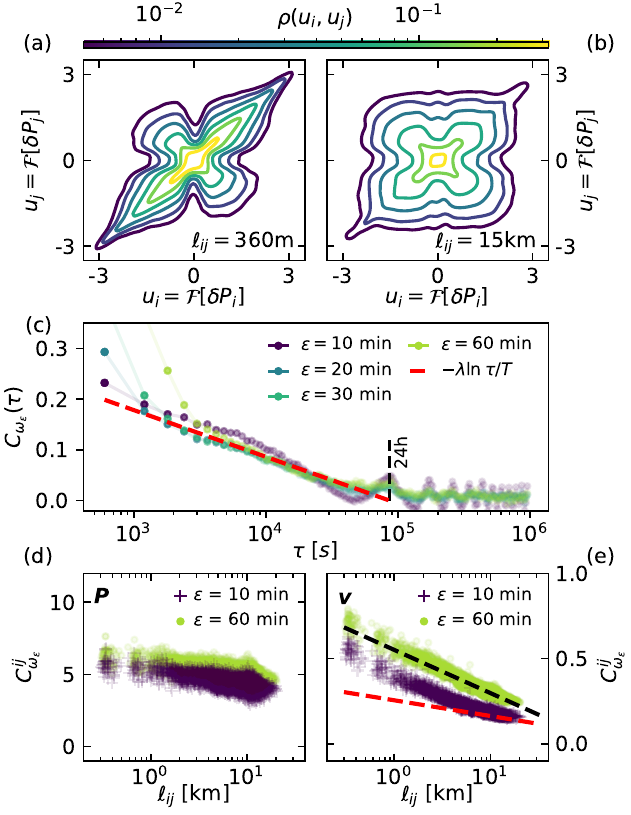}
    \caption{Copula analysis and spatial correlations. (a) and (b) Iso-density lines  of $(\delta_\tau P_i,\delta_\tau P_j)$'s Gaussian copulas $\rho(u_i,u_j)$, for (a) neighboring and (b) distanced turbines. (c) Wind speed autocovariance $C_{\omega_\varepsilon}$ versus lag $\tau$ for different $\varepsilon$ values. The red line is decaying logarithm of parameters $T=24\mathrm{h}$, and $\lambda = 0.04$~\cite{baile2010spatial}. (d) and (e) Covariances $C_{\omega_\varepsilon}^{ij}$ versus $\ell_{ij}$ (computed from Eq.~\eqref{eq:EmpiricalCovariances}) for $\varepsilon =10\mathrm{min, and\ } 60\mathrm{min}$, for $P$ in (d), and $v$ in (e). Black and red logarithmic curves respectively use $\lambda =0.10 $, $\xi = 200\mathrm{km}$ and $\lambda =0.04 $, $\xi = 600\mathrm{km}$ (similar to Ref.~\cite{baile2010spatial}).}  
    \label{fig:SpatialCorrelations}
\end{figure}

To eliminate the bias induced by the marginal distributions of power increments, we construct the associated Gaussian copulas by mapping each marginal onto a zero-mean, unit-variance normally distributed variable, \(u_i = \mathcal{F}[\delta_\tau P_i]\). The monotonic transformation \(\mathcal{F}\) is found to be approximately identical across turbines. The resulting copulas,
\[ \rho\left( u_i=\mathcal{F}[\delta_\tau P_i], u_j=\mathcal{F}[\delta_\tau P_j] \right) = \mathbb{P}(\delta_\tau P_i,\delta_\tau P_j), \]
are computed from Gaussian kernel density methods~\cite{Rosenblatt1956,Silverman2018} and are shown in Fig.~\ref{fig:SpatialCorrelations}(a) and (b) (see also Supplemental Material~V.~\bibnotemark[Supplemental]). They reveal a distinctly non-elliptical dependence structure, closely resembling that observed in joint return statistics in finance~\cite{chicheportiche2012notelliptical}, and the significance of which we now explain. At fixed \(\tau\), extreme events remain strongly correlated for all separations \(\ell_{ij}\), yielding a characteristic cross-shaped joint p.d.f., while for neighboring turbines their signs align preferentially, as evidenced by the pronounced elongation along the \(u_i=u_j\) direction. This joint alignment of signs and persistence of large amplitudes directly drives the excess intermittency, since coincident extreme fluctuations add coherently and dominate the aggregated statistics.\\

\textit{Magnitude correlations.---} To quantitatively measure the space-time range of extreme events correlations, we compute from our time-series a measure of local deviation, the \textit{log-volatility}  (or \textit{magnitude}) defined as:
\begin{equation}\omega_\varepsilon = \ln |\delta_\varepsilon X| +\Omega_\varepsilon,\label{eq:OmegaTimeDecay}\end{equation}
where the constant term $\Omega_\tau$ ensures $\langle \omega_\tau \rangle=0$.
Interestingly, that measure was introduced in Ref.~\cite{bacry2001multifractal} and used in Refs.~\cite{Vernede2015,Lakhal2025,baile2010spatial,muzy2010intermittency} as a complementary description of multifractal statistics. In fact, for a multifractal process of \textit{intermittency coefficient} $\lambda = -\zeta''_q(0)$, it was shown~\cite{bacry2001multifractal,muzy2010intermittency} that the covariance of its magnitude should decay following: 
\begin{equation}C_{\omega_\varepsilon}(\tau) = \langle \omega_\varepsilon(t)\omega_\varepsilon(t+\tau)\rangle =  -\lambda \ln{\tau/\mathcal{T}},\ \tau> \varepsilon
\label{eq:omegacorrelator}
\end{equation} with $\mathcal{T}$ an upper multifractal range, usually coinciding with upper inertial range $T$ ---although see Refs.~\cite{Vernede2015,Lakhal2025} for counterexamples. In Fig.~\ref{fig:SpatialCorrelations}(c), the autocovariance of wind speed magnitude $\omega_\tau^v$ computed from Eq.~\eqref{eq:omegacorrelator} expectedly shows logarithmic decay for $\tau>\varepsilon$, with $\mathcal{T} =T=24\mathrm{h}$, and $\lambda = 0.04$, identically reported in Ref.~\cite{baile2010spatial,muzy2010intermittency} for hourly sampled ($\varepsilon= 1\mathrm{h}$) Dutch and Corsican wind data. Note that fitting $H_q$ around $q\to 0$ from our wind speed data yields the same $\lambda = 0.04$ value. The excess correlations reported for $\tau<\varepsilon$, result from the coarse-graining introduced in the volatility computation. 

We extend our analysis to spatial correlations of the magnitude by defining and computing the pairwise covariance:
\begin{equation}
    C_{\omega_\varepsilon}^{ij} = \langle \omega^{i}_\varepsilon (t)\omega^{j}_\varepsilon (t)\rangle 
    \label{eq:EmpiricalCovariances}
\end{equation}
for all $80^2 = 1600$ pairs $(i,j)$ of magnitudes measures of $P$ and $v$, using $\varepsilon =$ 10 min and 60 min~\cite{baile2010spatial}. Figures.~\ref{fig:SpatialCorrelations}(d) and (e) show the decaying trend of these covariances as functions of turbine separation $\ell_{ij}$. For $\omega_P$, correlations were found to slightly decrease over turbine distance, an effect we ascribe to the presence of saturating and shutdown regimes. 

For \(\omega_v\), Fig.~\ref{fig:SpatialCorrelations}(d) shows that the spatial covariance organizes along logarithmically decaying curves as a function of turbine separation, of the form 
\begin{equation}
C_{\omega_\varepsilon}(\ell) = -\lambda \ln {\ell}/{\xi}.
\end{equation} 

For a coarse-graining timescale \(\varepsilon = 1\,\mathrm{h}\), a fit (black curve in Fig.~\ref{fig:SpatialCorrelations}(d)) yields \(\xi = 200 \pm 50\,\mathrm{km} \simeq L/2\), slightly below the forcing scale associated with diurnal variability, and \(\lambda = 0.10 \pm 0.01\), significantly larger than the value inferred from the temporal correlations of \(\omega_\varepsilon\) in the \(\tau > \varepsilon\) regime.

In reality, to evidence a universal correlation regime for $\omega_v$, the sweeping approximation yields a condition similar to $\tau>\varepsilon$ found for the auto covariance analysis: \(\ell > \varepsilon\,u_{\mathrm{RMS}}\), which for $\varepsilon=60\mathrm{min}$ leads to quasi-maximum turbine separation \(\ell = 18\mathrm{km}\sim D\). As a result, logarithmic fits obtained for larger values of \(\varepsilon\) will display excess correlations and coarse-graining dependent fitting parameters.

For \(\varepsilon = 10\,\mathrm{min}\), the sweeping condition reduces to \(\ell > 3\,\mathrm{km}\). In this case, Fig.~\ref{fig:SpatialCorrelations}(d) reveals a crossover from excess short-range correlations to a logarithmic decay consistent with the parameters \(\lambda = 0.04\) and \(\xi = 600\,\mathrm{km}\) (black curve) reported in Ref.~\cite{baile2010spatial} for \(\ell > \varepsilon\), although data points around \(\ell \sim 20\,\mathrm{km}\) remain systematically above the fit. This value of \(\xi\) is consistent with previous estimates of correlation lengths for instantaneous (\(\varepsilon \to 0\)) power fluctuations across wind plants in Texas (\(L \approx 300\,\mathrm{km}\))~\cite{katzenstein2010variability}, as well as with the estimate \(L \approx 400\,\mathrm{km}\) obtained from the sweeping approximation.\\

Ultimately, in the context of wind power, our results show that amplitude correlations extend over distances comparable to, and well beyond, the physical size of the wind farm. We further identify a clear crossover from excess short-range correlations induced by signal filtering to a surface layer boundary turbulent regime, consistent with previously reported statistics~\cite{baile2010spatial}. As a consequence, spatial aggregation does not suppress extremes: rather, extreme fluctuations combine coherently, allowing the farm power output to retain and even amplify its non-Gaussian character.

\section{Conclusion \label{sec:Conclusion}}

\textit{Results.---} We conclude by summarizing our main findings. We have shown that wind turbines inherit the scale-invariant and non-Gaussian statistics of atmospheric wind-speed fluctuations. At the farm level, however, the aggregated power output departs from single-turbine behavior, by displaying steeper spectral decay and more intermittent statistics. These features were investigated through a combination of cross-correlation analyses and nonlinear multivariate descriptions, applied consistently to both wind speed and power signals.

We demonstrated the existence of robust and universal cross-correlations between turbine outputs. These correlations show a transition from short-time decoherence to turbulence-driven scaling regime, controlled by a coherence time \(\tau_{ij}\) directly related to turbine pair-distance \(\ell_{ij}\) In particular, we found the sublinear relation \(\tau_{ij}\sim \ell_{ij}^\beta\) with \(\beta<1\), indicating the stronger retardation of smaller scale structures in the farm.

By using Gaussian copula and autocorrelation analyses, we further showed that wind-power fluctuations exhibited persistent correlations in both amplitude and sign across turbines. In particular, and consistently with mesoscale boundary layer fluctuations, amplitude correlations extend well beyond the physical size of the farm (\(\sim 20\,\mathrm{km}\)). As a result, extreme events in neighboring turbines align in jump directions, leading to coherent summation of power bursts, thus reinforcing intermittent behavior in the aggregated output.

We systematically extended our analysis to ideal turbines with unbounded rated power (\(P_{\mathrm{sat}} = +\infty\), \(P \propto v^3\)). In this limit, the scaling of individual turbine production was found to deviate from the Kolmogorov scaling of wind speed, a behavior we attribute to the auto-regressive nature of wind-speed fluctuations. Despite the absence of saturation effects, the aggregated power in this idealized setting exhibits the same increase in persistency (jump sign correlations) and intermittency reported for real turbines. This confirms that non-linear response regimes and  turbulence-induced correlations are both independently responsible for the robustness of non-Gaussian
features under wind power aggregation.\\

\textit{Perspectives.---}
Our results rely on the scale-invariant properties of atmospheric wind turbulence; we therefore expect the reported findings to remain valid, and to be further refined, at higher temporal resolutions. In particular a better identification of causal wind structures, boundary-layer variability, and $\varepsilon$-independent magnitude correlations would be beneficial.

For the resolution considered $\Delta t = 10\text{min}$, the mapping wind speed/power introduced by the response function $P = f(v)$  suggests practical routes for forecasting intraday power fluctuations. One option would be to enrich multivariate wind speed models with a spatial prior in order to improve short-time predictions, as explored in Refs.~\cite{behnken2020,calif2014}, where Fokker--Planck-based predictors for wind speed and power were developed—albeit relying solely on local-in-space measurements. In this context, the long-range magnitude correlations identified here, and previously reported in the literature~\cite{baile2010spatial}, provide a valuable basis for wind-speed volatility forecasting, in close analogy with methods developed for asset-price dynamics~\cite{sattaroff2023,Muzy01012001,bacry2001modelling}.

More broadly, the framework developed in this work provides quantitative tools to assess wind-farm sensitivity to extreme events and to evaluate their impact on electrical infrastructure. In a way similar to the phenomenological description of financial markets~\cite{garnier2021new,bouchaud2001leverage,aubrun2024riding}, we expect that the continued uncovering and characterization of the nonlinear correlation structure of wind power fluctuations will provide actionable insights for local buffering and control strategies.

While the large-scale transition toward sustainability will rely on coordinated policies and cross-system collaboration~\cite{veers2019grand,IEA_Task25_2018}, we expect our results, together with recent advances in renewable variability, to provide quantitative benchmarks to inform and calibrate future policy and operational decisions.

\begin{acknowledgments}  The authors thank C. P. Connaughton, A. Pumir, J. Liu, J.-P. Bouchaud, C. Aubrun, J. Garnier-Brun, and members of the Econophysics \& Complex Systems Research Chair (École polytechnique) for scientific discussions. SL was supported by a Japan Society for the Promotion of Science (JSPS) Postdoctoral Fellowship (grant no. P24714). MB was supported by JSPS KAKENHI (Grant No. 24KF0079). The data used in this work was obtained from Scout Clean Energy, Boulder CO.\end{acknowledgments}

\bibliography{Bibliography}

\end{document}